\documentclass{article}
\usepackage[a4paper, total={6.3in, 8.2in}]{geometry}
\usepackage[utf8]{inputenc}
\usepackage[T1]{fontenc}

\usepackage{graphicx}%
\usepackage{multirow}%
\usepackage{amsmath,amssymb,amsfonts}%
\usepackage{amsthm}%
\usepackage{mathrsfs}%
\usepackage[title]{appendix}%
\usepackage{xcolor}%
\usepackage{textcomp}%
\usepackage{manyfoot}%
\usepackage{booktabs}%
\usepackage{algorithm}%
\usepackage{algorithmicx}%
\usepackage{algpseudocode}%
\usepackage{listings}%

\usepackage[separate-uncertainty=true, per-mode=repeated-symbol]{siunitx}

\usepackage{hyperref}
\usepackage{here}
\usepackage{pifont}
\usepackage{subcaption}

\usepackage[
    style=nature,
    backend=biber,
    doi=true,
    isbn=false,
    url=false,
    eprint=false,
    maxnames=3,
    minnames=1,
    giveninits=true,
    terseinits=true
]{biblatex}

\AtEveryBibitem{%
    \clearfield{language}%
    \clearfield{issn}%
    \clearfield{note}%
}

\usepackage{comment}

\usepackage{xargs}

\def\commentType{1}
\ifnum\commentType=0
    \newcommandx{\customComment}[3]{}
    \newcommandx{\customTODO}[3]{}
\fi
\ifnum\commentType=1
    \newcommandx{\customComment}[3]{\textcolor{#2}{\textsl{#1: #3}}}
    \newcommandx{\customTODO}[3]{\textcolor{#2}{\textsl{#1: #3}}}
\fi
\ifnum\commentType=2
    \usepackage{pdfcomment}
    \newcommandx{\customComment}[3]{\pdfcomment[icon=Comment,opacity=0.5,color=#2,author=#1]{#3}}
    \newcommandx{\customTODO}[3]{\pdfcomment[icon=Note,opacity=0.5,color=#2,author=#1]{#3}}
\fi
\ifnum\commentType=3
    \usepackage{pdfcomment} 
    \usepackage{todonotes} 
    \usepackage{silence}
    \newcommandx{\customComment}[3]{\todo[color=#2!40,size=\small]{\textbf{#1:} #3}}
    \newcommandx{\customTODO}[3]{\todo[color=#2!40,size=\small]{\textbf{#1:} #3}}
\fi

\usepackage{ulem}
\newcommandx{\added}[1]{#1}
\newcommandx{\removed}[1]{}
\newcommandx{\replaced}[2]{\removed{#1}\added{#2}}

\newcommandx{\radded}[1]{#1}
\newcommandx{\rremoved}[1]{}

\newcommandx{\rradded}[1]{#1}
\newcommandx{\rrremoved}[1]{}
\newcommandx{\rrreplaced}[2]{\rradded{#2}}

\usepackage{xr-hyper}
\newcommand{\secnameref}[1]{Section \hyperref[#1]{\nameref*{#1}}}
\newcommand{\suppsecnameref}[1]{Supplement \hyperref[#1]{\nameref*{#1}}}

\begin{document}
\title{Overprinting with \\Tomographic Volumetric Additive Manufacturing}

\author{
  Felix Wechsler\textsuperscript{1,*},
  Viola Sgarminato\textsuperscript{1,2},
  Riccardo Rizzo\textsuperscript{1},\\
  Baptiste Nicolet\textsuperscript{3,4},
  Wenzel Jakob\textsuperscript{3,*},
  Christophe Moser\textsuperscript{1,*}
  \\[2ex] 
  \textsuperscript{1}Laboratory of Applied Photonics Devices,\\École polytechnique fédérale de Lausanne (EPFL), Lausanne, Switzerland\\  
   \textsuperscript{2}BIOINSIDE Lab, Department of Mechanical and Aerospace Engineering,\\Politecnico di Torino, Italy\\
  \textsuperscript{3}Realistic Graphics Lab,\\École polytechnique fédérale de Lausanne (EPFL), Lausanne, Switzerland\\
  \textsuperscript{4}NVIDIA, Zürich, Switzerland
  \\[2ex] 
  \textsuperscript{*}Corresponding authors:\\ \texttt{overprinting@felixwechsler.science},\\ \texttt{wenzel.jakob@epfl.ch},\\ \texttt{christophe.moser@epfl.ch}
}

\maketitle
\begin{abstract}
    \textcolor{red}{\large \textbf{You are viewing a preprint of this peer-reviewed manuscript:\\ \url{https://doi.org/10.1038/s41467-026-73477-3}}}\\

    Tomographic Volumetric Additive Manufacturing (TVAM) is a light-based 3D printing technique capable of producing centimeter-scale objects within seconds. A key challenge lies in the calculation of tomographic projection patterns
    under non-standard conditions, such as the presence of occlusions and materials with diverse optical properties, including varying refractive indices or scattering surfaces.\\
    This work demonstrates a broad range of overprinting scenarios, where new structures are directly printed onto or around pre-existing components made from different materials.
    Our simulations and experimental verifications perform overprinting of absorbing, refracting, reflecting and scattering elements in
    both round and square vials.\\
    All scenarios are optimized with our differentiable, physically based ray-optics approach using the open-source Dr.TVAM framework, delivering high-quality projections for both laser- and LED-based illuminations within minutes and lower-quality projections within seconds, exceeding existing open-source solutions in speed, flexibility, and quality.
\end{abstract}

\newpage
\section*{Introduction}\label{sec1}
    Tomographic volumetric additive manufacturing (TVAM) \cite{kelly2019volumetric, 
    bernal2019volumetric, Loterie_Delrot_Moser_2020} is a rapid light-based 3D printing technique that enables 
    manufacturing of centimeter-scale structures in seconds. Pre-calculated 2D 
    patterns are projected onto a rotating vial filled with a photosensitive 
    (single-photon absorption) resin. The patterns propagate into the resin and 
    deposit a 3D energy dose over time. Once the accumulated energy dose \removed{exceeds}
    exceeds a threshold, local network formation occurs \cite{gibson2015vatphotopolymerization}.

    Due to the similarity between TVAM and computed tomography (CT), the Radon 
    transform and its adjoint (the backprojection) can be employed as a physical 
    forward model for light propagation. However, these models neglect the 
    intensity loss of light as it propagates through the medium, making the 
    attenuated Radon transform a more appropriate model \cite{FNatterer_2001, 
    kelly2019volumetric}. In the early stages, light patterns for TVAM were 
    calculated using filtered backprojection. However, this approach resulted in 
    physically impossible negative projection intensities. While these values are easily clipped to zero, doing so leads to an energy mismatch that causes serious artifacts into the final prints. Rackson et al. \cite{Rackson_Champley_Toombs_Fong_Bansal_Taylor_Shusteff_McLeod_2021} 
    utilized the thresholding behavior to develop an iterative algorithm that 
    ensures object regions receive more dose than an upper threshold while 
    void regions remain below a lower threshold. This concept was transformed into a loss function that was then 
    optimized using a gradient descent-based optimizer \cite{Wechsler_Gigli_Madrid-Wolff_Moser_2024, 
    refId0}. A more generalized version of a similar loss function has been introduced by Li et al. \cite{Li_Toombs_Taylor_Wallin_2024}.
    
    Moreover, the rotating vial was initially placed in an index-matching bath to 
    approximately eliminate light refractions at the air-to-vial-to-resin interfaces. 
    Later, this limitation was addressed by correcting the patterns for refraction 
    \cite{loterie2022method, Orth_Sampson_Ting_Boisvert_Paquet_2021}, allowing 
    vials to be used in air. However, this post-processing is only an approximation, 
    as it cannot be accurately combined with the correct attenuation of the rays 
    within the vial. Consequently, more powerful frameworks based on ray tracing 
    schemes have been proposed \cite{webber2023versatile, 
    Nicolet_Wechsler_Madrid-Wolff_Moser_Jakob_2024}. In these works, rays are 
    traced through a voxel grid, allowing for correct energy deposition and 
    refraction accounting.
    
    TVAM has proven to be extremely flexible and has been utilized for a wide range 
    of materials and printing scenarios \cite{Madrid-Wolff_Toombs_Rizzo}. 
    Particularly, modeling light scattering has been successful in printing in 
    media with scattering particles \cite{Nicolet_Wechsler_Madrid-Wolff_Moser_Jakob_2024}, such as bio-resins \cite{madrid2022controlling}. 
    Early TVAM setups employed laser diodes (referred to as LaserTVAM), which resulted in striation artifacts \cite{kewitsch1996selffocusingPhotopolymerization, Rackson:22}. 
    By using LED illumination for TVAM (LEDTVAM), these striations were largely 
    mitigated, enabling the production of optical elements with smooth surfaces 
    \cite{Webber_Zhang_Sampson_Picard_Lacelle_Paquet_Boisvert_Orth_2024}.
    As the chemical interactions affect the final print, previous work modeled  the physical and chemical processes of TVAM \cite{weisgraber2023virtual}
    and corrected for effects such as diffusion of inhibitors \cite{Orth_Webber_Zhang_Sampson_de}. Also holographic projection schemes with 
    digital micromirror devices have been used to increase the laser light engine efficiency \cite{holotvam}.
    
    Due to the rotation of the vial \added{the printing projector has access to different sides of existing structures. Hence,} overprinting of pre-existing structures has emerged 
    as a potential application of TVAM. 
    Indeed, other 3D printing techniques are 
    often limited because they build the 3D structure from one side and therefore 
    cannot print effectively around occlusions. 
    This capability has been demonstrated 
    in the printing of a handle on a metal screwdriver \cite{kelly2019volumetric}. 
    However, this work assumed the metal rod to be fully absorptive, which is not 
    appropriate for many surfaces. 
    Later, more complex geometries were overprinted 
    over existing structures \cite{Chansoria_ruetsche_Wang}, although in this work 
    the existing structure was completely disregarded in the pattern optimization. 
    Similarly, a gelatin methacrylate (Gel-MA) hydrogel was successfully overprinted 
    over an endoskeletal system \cite{Darkes_Burkey_Shepherd_2024}.
    
    Recently, a more complex method called GRACE (Generative, Adaptive, Context-Aware 3D Printing)  \cite{Florczak_Groessbacher_Ribezzi_Longoni_Gueye_Grandidier_Malda_Levato_2024} was introduced. 
    Using a light sheet imaging system, the authors detected spheres and overprinted 
    connected channels over these structures. However, the light transport model still seems to be limited to fully transparent or fully blocking occlusions.
 
    In recent years, several studies have demonstrated the potential of TVAM as a bioprinting technique capable of fabricating cellularized constructs with diverse geometries, sizes, and internal cavities that closely mimic human physiology \cite{https://doi.org/10.1002/adma.201904209,Sgarminato_2024}. Indeed, TVAM is particularly well-suited for printing hollow 3D structures, as it does not require sacrificial support materials. The integration of bioprinting with microfluidics has gained increasing attention, as it enables the fabrication of anatomically relevant 2.5D and 3D structures within dynamic and physiologically realistic environments \cite{D4BM01354A}. 
    For example, TVAM has also been explored for the fabrication of perfusable constructs with potential for microfluidic systems and organ-on-chip \cite{RizzoMultiscale, AmeliaScienceAdvances}. However, efforts to date have revolved around standard, multi-step processes that include printing in round vials, recovering and postprocessing of the printed object, and finally assembly of the multicomponent microfluidic chamber. In addition to being tedious and inefficient, this procedure significantly increases the risk of leakage and contamination. Similar issues are intrinsic to alternative approaches such as embedded printing and digital light processing. 
    
    Recently, Nicolet et al. \cite{Nicolet_Wechsler_Madrid-Wolff_Moser_Jakob_2024} introduced a physically-based differentiable simulation software for TVAM, named Dr.TVAM, built on the open-source differentiable renderer Mitsuba 3 \cite{Mitsuba3}. 
    Since it is based on a general rendering system, Dr.TVAM enables  integration of other components or occlusions in the printing vial, and supports customizable printing geometries. 
    Its physically-based optical simulation of the TVAM process allows to account for various effects such as scattering, absorption, arbitrary reflections and refractions. 
    
    In this work we introduce overprinting scenarios and highlight the versatility of our framework Dr. TVAM across various contexts and applications. Our simulations and experiments build on the versatility of this platform for pattern optimization.
    
    First, we lay the foundation for a new, streamlined workflow of the biofabrication of microfluidic chips. By accounting for unconventional vial geometry (square cuvette) and occluding elements (inlets and outlets), Dr.TVAM enables overprinting of microfluidic networks into preassembled 3D chips, thus opening to a new generation of on-chip technologies. Based on this workflow, we also overprinted small glass spheres at arbitrary locations by detecting them and hereby showing context-aware fabrication of microfluidic chips. Next, we print a simple gear on a reflective rod and demonstrate that more complex light modeling outperforms existing solutions.
    \replaced{Finally, we overprint lenses on existing elements such as glass tubes and LEDs.}{Finally, we overprint a mini projector on a LED.} Furthermore, its ability to handle non-telecentric configurations makes Dr.TVAM, to the best of our knowledge, the only \replaced{publicly}{open-source} available software for the simulation and optimization of LED-based TVAM systems.
    
    All source code, 3D meshes and configuration scripts are released for reproducibility.

\section*{Results}\label{sec2}

    \subsection*{Perfusion system for bio-applications}
        \label{results:perfusion}

\begin{figure}[H]
            \centering
            \includegraphics[width=1\textwidth]{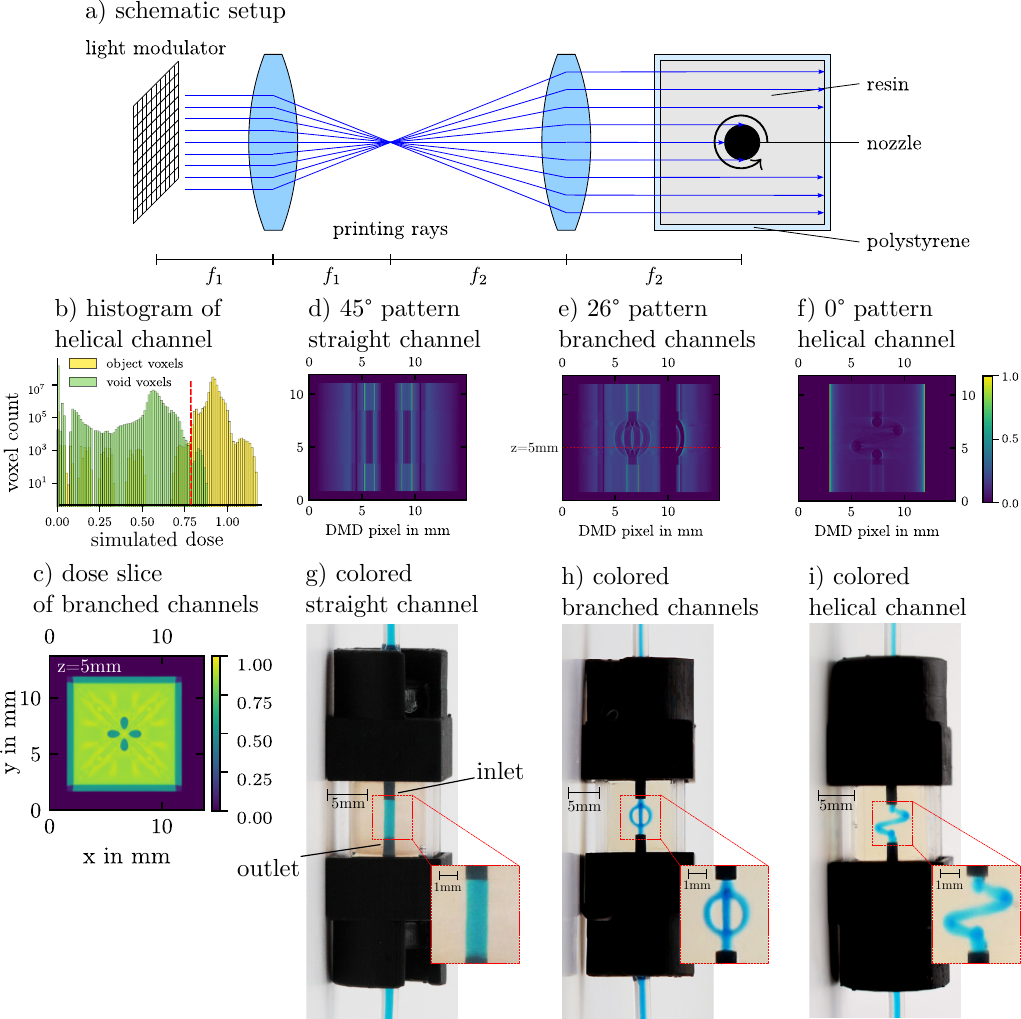}
                \caption{Fabrication of perfusable microfluidic channels in a pre-assembled, square cuvette. 
                a) Schematic setup with projections into a square cuvette. 
                b) Voxel \replaced{intensity}{dose} histogram from the simulation. 
                c) A slice of the cumulative \replaced{intensity}{dose} map. 
                d, e, f) Example projection patterns calculated by Dr.TVAM for different angles. The pattern in d) clearly shows the refractive effect of the square cuvette walls. 
                g, h, i) Final printed microfluidic channels with different geometries inside the sealed cuvette. }
                \label{fig:perfusion}
        \end{figure}

        To demonstrate the capability of Dr.TVAM in fabricating microfluidic chips within flat, imaging-compatible vials using a biocompatible resin, we designed a custom platform. This platform consists of a transparent chamber made from a cut \SI{1}{\centi\meter} polystyrene \added{square-shaped} cuvette (see \autoref{fig:perfusion}). Stereolithography (SLA)-printed adapters featuring inlets, outlets, barbed connectors for tubing, and circular supports for mounting to a rotational stage were press-fitted onto both ends, creating a leak-proof system. 
        The widely used biocompatible photopolymer gelatin methacryloyl (Gel-MA) with LAP as the photoinitiator was used for printing (see \secnameref{sec:gelmaresin} for details).\\
        In this overprinting scenario, the adapters were fabricated from black resin, so the inlet and outlet occlusions were assumed to be fully light-absorbing. Various perfusable models, including a simple straight channel, a spiral, and a branched geometry, were optimized using Dr.TVAM, accounting for the square container  and the presence of these occluding elements. 
        \added{In TVAM typically round vials are used to contain the resin but since our framework is very general, arbitrary container shapes are possible.}
        This experiment was performed with the laser-based TVAM setup (LaserTVAM) described in \secnameref{sec:lasertvam}. A schematic of the LaserTVAM setup is shown in \autoref{fig:perfusion}a.
        The optimization results are shown in \autoref{fig:perfusion}b  and c, which display the \replaced{intensity}{dose} histogram and a slice of the final \replaced{intensity}{dose} distribution after projection of the patterns, respectively.

        The histogram plots the voxel intensities for both object and void regions.
        Note that the voxel count is on a logarithmic scale.
        \added{Histograms plots are an effective way to judge if there is a good separation between void and object exposure \cite{Rackson_Champley_Toombs_Fong_Bansal_Taylor_Shusteff_McLeod_2021}}.
        Although the \replaced{intensity}{dose} in \autoref{fig:perfusion}c is not perfectly homogeneous, applying a threshold (\added{indicated by a dashed red line in the histogram}) yields a best Intersection over Union (IoU) score of 0.9972, which is close to the maximum possible value \added{of 1}.
        Example projection patterns are shown in \autoref{fig:perfusion}d, e, and f. For instance, \autoref{fig:perfusion}d shows the pattern for a projection angle of \SI{45}{\degree}.
        Here, light is refracted through both the left and right sides of the cuvette, highlighting the influence of its square geometry. Furthermore, the optimization restricts light from passing through the corners of the cuvette, as they are frosted and would cause unwanted scattering.

        The final prints for three different target shapes are shown in \autoref{fig:perfusion}g, h), and i).
        \added{The designed channel diameters are \SI{0.95}{\milli\meter}, \SI{0.7}{\milli\meter} and \SI{0.95}{\milli\meter} respectively.
    Experimentally, we measure \SI{1.15\pm0.1}{\milli\meter}, \SI{0.5\pm0.1}{\milli\meter} and \SI{0.6\pm0.1}{\milli\meter}. The deviations can be explained by over- or underpolymerization, swelling of the material but also diffusion effects of the dye.}

        Videos of the perfusion can be found in the supplementing material (Supplementary Movie 1, Supplementary Movie 2, Supplementary Movie 3).
        This \replaced{flat}{non-curved} \added{container} geometry simplifies imaging, and when used in pre-assembled chips, as in this case, it also allows for the container to remain sealed post-fabrication, thereby maintaining sterile conditions.
        In summary, using  Dr.TVAM we successfully generated optimized patterns for an unconventional square vial geometry with internal occlusions, enabling the fabrication of perfusable microfluidic features within customizable pre-assembled chips.

    \subsection*{Context-aware perfusion system}
            \label{results:channels}

\begin{figure}[h]
            \centering
            \includegraphics[width=\textwidth]{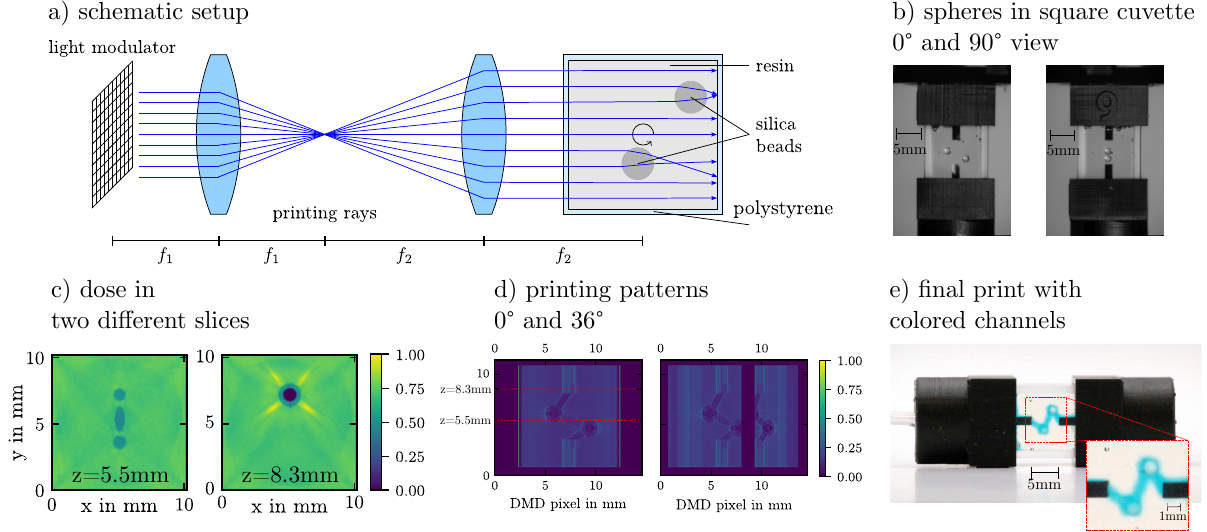}
            \caption{Rapid fabrication of a context-aware microfluidic chip platform. 
                a) Schematic setup of the experiment. 
                b) Images of the spheres embedded in Gel-MA. 
                \removed{c) Histogram of the optimization results.}
                \replaced{d)}{c)} Cumulative \replaced{intensity}{dose} distribution across the printing area. 
                \replaced{e)}{d)} Two example projection patterns used for printing. 
                \replaced{f)}{e)} Final print showcasing colored channels around the embedded spheres.}
            \label{fig:balls}
        \end{figure}

        In this next example, we used the microfluidic chip platform described previously to showcase the capability of Dr.TVAM to rapidly generate \removed{on-the-fly} projection patterns for case-specific, context-aware printing (\autoref{fig:balls}).

        The process began by filling the square chamber with Gel-MA and positioning two \SI{1}{\milli\meter} glass spheres between the nozzles see \autoref{fig:balls}a and b. \added{Since Gel-MA is solid at room temperature, we did not observe any movement because of gravity (see also \secnameref{sec:sphereover}).} 
        The spheres possess a different refractive index and refract the light, as indicated in \autoref{fig:balls}a.
        A camera-based system captured the coordinates of both spheres from two orthogonal views (\SI{0}{\degree} and \SI{90}{\degree}), as seen in \autoref{fig:balls}b . The flat geometry of the cuvette allows for the straightforward determination of the spheres' 3D positions from these two views. By measuring the pixel positions, we extracted the real-world coordinates in under a minute.\\
        Based on these positions, we generated a target geometry and optimized the printing patterns, treating the glass spheres as reflective and refractive occlusions. The quality of the optical simulation was reduced to achieve a fast pattern optimization time \added{(see details in \secnameref{sec:sphereover})} of approximately \SI{30}{\second}. The generated patterns defined hollow straight channels (inner diameter \SI{0.7}{\milli\meter}) connecting each sphere to the inlet and outlet of the perfusion system. Additionally, spherical cavities (inner diameter \SI{1.7}{\milli\meter}) were printed around the spheres to secure their position while enabling liquid perfusion.
        The \replaced{intensity}{dose} distribution of two different vertical slices in \autoref{fig:balls}\replaced{d}{c}. Although the \replaced{intensity}{dose} distribution is not perfectly homogeneous, the highest IoU of 0.9951 indicates successful optimization. Two example patterns are displayed in  \autoref{fig:balls}\replaced{e}{d}. 
        
        To evaluate the performance of the overprinting algorithm, the final print was perfused with a blue food dye solution. As shown in \autoref{fig:balls}\replaced{f}{e}, this approach enables the rapid fabrication of functional channels around millimeter-scale objects with arbitrary spatial positions within the chamber.

    \subsection*{Gear on metal rod}
    \label{results:gears_rod}
        As previously reported, Kelly et al. \cite{kelly2019volumetric} successfully printed a handle over an existing metallic screwdriver rod. In their work, however, the metal rod was assumed to be fully light-absorbing. This approximation, while not physically accurate, can be sufficient in specific cases. However, common metal parts like polished steel are highly reflective. In such cases, a significant amount of light is scattered or reflected, which must be taken into account. 
        \rremoved{More details how we describe such a rod in our model, are available in }
        \begin{figure}[h]
            \centering
            \includegraphics[width=1\textwidth]{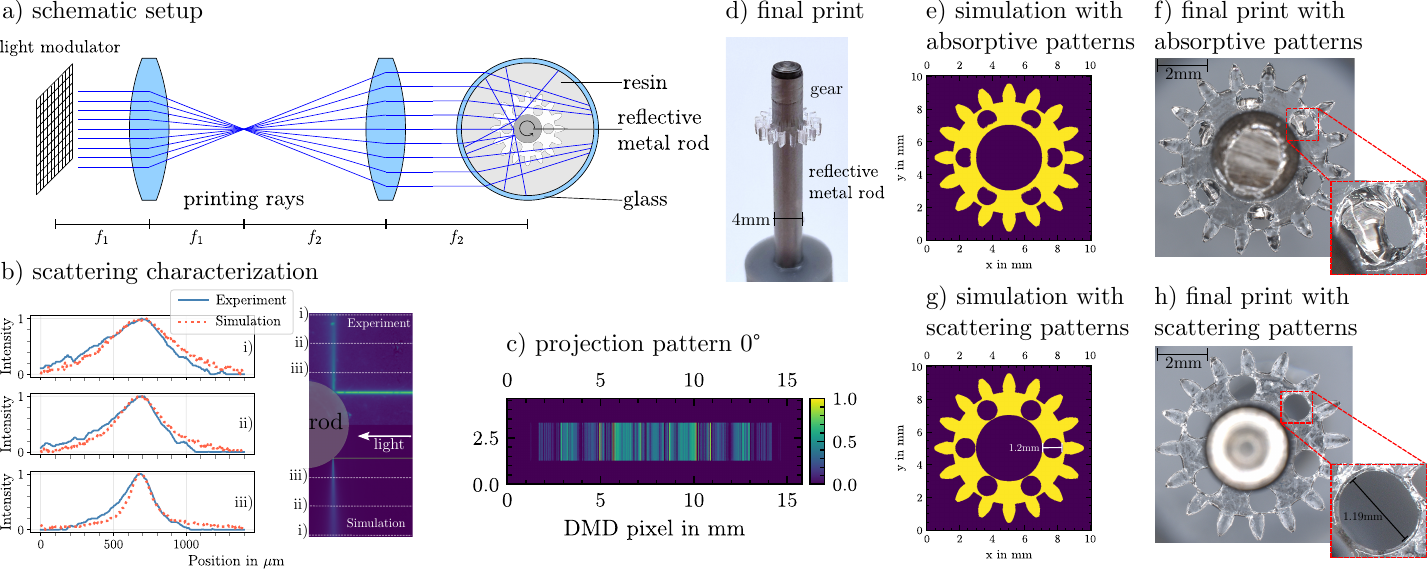}
            \caption{\radded{Comparison of printing results assuming an absorptive vs. a reflective rod. 
                a) Schematic of the experimental setup. 
                b) Characterization of scattering behaviour results in $\alpha=0.03$ used for the Beckmann distribution.
                c) An example of an optimized light pattern. 
                d) The final printed part.
                \textbf{Top row:} Absorptive rod assumption. 
                e) simulated print with absorbing patterns projected onto a reflecting rod and f) final print.
                \textbf{Bottom row:} Scattering rod assumption. 
                g) simulated print with scattering patterns projected onto a reflecting rod and h) final print.}}
            \label{fig:gears}
        \end{figure}
  
        To demonstrate our approach experimentally, we printed a gear featuring teeth and circular holes onto a polished steel rod using an acrylate resin (see \secnameref{sec:acrylate_resin} for details). 
        A 3D SLA-printed cap was used to position the metal rod in the center of the glass vial. We optimized the light patterns for two scenarios. The first assumed a perfectly absorbing rod, as in previous work. The second used a more realistic model of a rough, light-scattering surface.
        This experiment was performed with  LaserTVAM (see \autoref{fig:gears}a).
        \radded{To characterize the scattering properties of the metal rod, we immersed it into a fluorescent water bath and analyzed the scattered light path from which we can infer a scattering parameter of the Beckmann distribution of $\alpha=0.03$. The simulated and experimental light paths and three cross-sections are shown in \autoref{fig:gears}b) each. See more details in \secnameref{meth:gears}}.
        \\
        First, we optimized the patterns for a fully absorbing rod.
        The \rremoved{cumulative dose}{final print} in the object space after projecting all patterns is displayed in \autoref{fig:gears}\rremoved{b}\radded{e)}. The resulting print using these patterns is shown in \autoref{fig:gears}\rremoved{c}\radded{f)}. In both the simulation and the experiment, the inner holes of the gear are over-polymerized and incorrectly formed. This result is expected, as these patterns do not account for the \replaced{energy}{light} scattered from the metal rod. \\
        In contrast, when we optimize the patterns in Dr.TVAM assuming a reflective metal rod, we achieve better results.
        One of the optimized patterns is displayed in \autoref{fig:gears}\rremoved{d}\radded{c)}. 
        \radded{The simulated print in \autoref{fig:gears}g) shows significant improvements over the absorptive case.}
        The gear's holes are well-preserved and by selecting the optimal threshold this optimized projection set reaches a simulated IoU of \rremoved{0.998}\radded{0.997} whereas ignoring the scattered light reduces the IoU to \rremoved{$0.958$}\radded{0.942}. 
        The final print is shown in \autoref{fig:gears}\rremoved{e}\radded{d)} and \rremoved{g}\radded{h)}. For both scenarios, we conducted a series of experiments with varying exposure levels and selected the print with the highest visual fidelity \radded{(see \suppsecnameref{supp:sec:exposure_series} for the experimental results at different power levels)}.
        \added{The adhesion of the gear to the rod is relatively high as shown in a demonstration video in the supplementing material (S11).}
        In summary, by accounting for light scattered from the metal rod, Dr.TVAM preserves features close to the rod's surface. In contrast, a model that assumes a fully absorptive rod cannot achieve the same printing fidelity in either simulation or experiment.

     \subsection*{Optical lens with \replaced{engraving}{symbol} on LED}
      \label{results:led}
        In this experiment, we print a lens and a lens holder with \replaced{engravings}{a symbol} directly onto a red LED using an acrylate resin. 
        \begin{figure}[h]
            \centering
            \includegraphics[width=\textwidth]{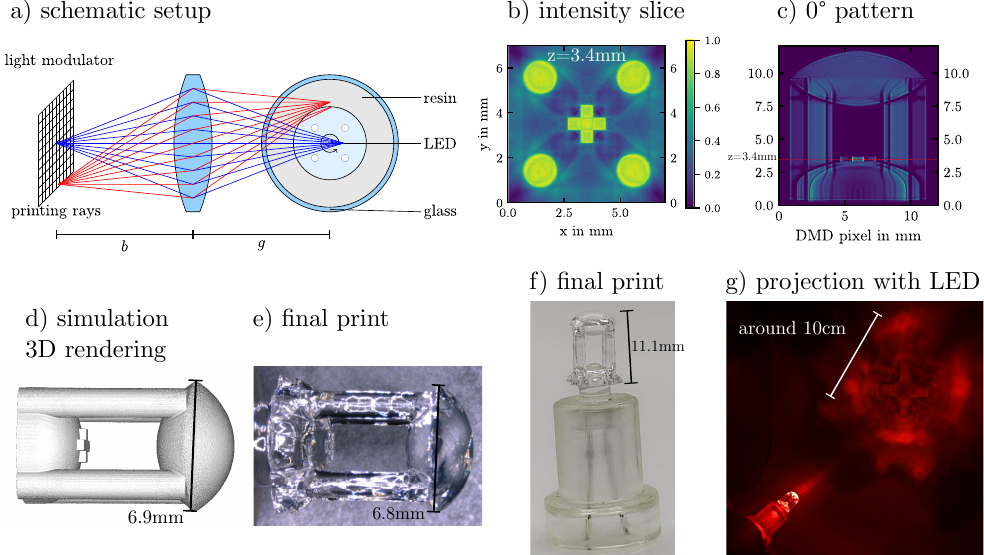}
            \caption{Printing a lens with \replaced{engravings}{a symbol} directly onto a red LED using LEDTVAM. 
                a) Schematic of the general setup. 
                \removed{b) Voxel \replaced{intensity}{dose} histogram from the simulation}
                \replaced{c}{b}) A slice of the dose after projection of all patterns. 
                \replaced{d}{c}) An example of a single projection pattern. 
                \replaced{e}{d}) A rendering of the overprinted LED and elements.
                \replaced{f}{e}) The final printed lens on the LED.
                \added{f}) The print on the LED and on the \added{holder}.
                g) The final image when the overprinted LED is turned on, projecting the $\mathord{\text{\ding{58}}}$ symbol onto a screen.}
            \label{fig:LED}
        \end{figure}
        To achieve an optically smooth surface, we use a non-telecentric, LED-based TVAM system (LEDTVAM), which is similar to the one described by Webber et al. \cite{Webber_Zhang_Sampson_Picard_Lacelle_Paquet_Boisvert_Orth_2024} and detailed in  \secnameref{sec:LEDTVAM}.
        The light paths in this scenario are complex, involving multiple refractions, reflections, and absorptions at various surfaces and within different media (air, glass, resin, and the LED itself). Our framework, Dr.TVAM, calculates these interactions based on the refractive indices, geometry, and material parameters. The general setup is shown in \autoref{fig:LED}a , with an LED positioned in the center of the vial. Unlike the collimated rays in the LaserTVAM, each pixel in the LEDTVAM projects a finite cone of light into the resin. Consequently, the resolution of the LEDTVAM is expected to be lower, as its depth of field is dependent on the specific aperture and imaging parameters. 
        Based on Mitsuba, we introduce these non-telecentric setups to Dr.TVAM and make it publicly available.\\
        \removed{Optimizing the patterns for this setup yields the well-separated \replaced{intensity}{dose} histogram shown in Figure 4b, although the separation is less distinct than in the LaserTVAM's histograms.}
        As shown in \autoref{fig:LED}\replaced{c}{b}, the optimization produces a sharp $\mathord{\text{\ding{58}}}$ symbol in the center of the cumulative \replaced{intensity}{dose} plot.  An example of a single projection pattern is shown in \autoref{fig:LED}\replaced{d}{c}.\\
        \added{The theoretical print is shown in \autoref{fig:LED}d}.
        The resulting \added{experimental} print is shown in \autoref{fig:LED}\replaced{f}{e}. The lens was printed successfully on top of the LED, and the $\mathord{\text{\ding{58}}}$ symbol is clearly visible on the coating. 
        \added{To place the actual LED inside the resin an SLA-printed cap was press-fit into the vial.}
        \removed{A schematic 3D rendering of the scenario can be seen in Figure 4e.}
        The purpose of this lens is to reimage the $\mathord{\text{\ding{58}}}$ symbol; when the LED is turned on, it projects the symbol onto a screen (\autoref{fig:LED}g).\\
        \removed{The print required over-polymerization to ensure structural stability, as lower exposure levels resulted in the structure collapsing or detaching from the LED. Although this caused defects on the LED coating, it did not significantly affect the functionality of this print.}

\section*{Discussion}\label{sec12}
    
  Dr. TVAM, our open-source framework \cite{Nicolet_Wechsler_Madrid-Wolff_Moser_Jakob_2024}, has the potential to significantly increase the fidelity of light-based additive manufacturing in non-standard optical conditions. By introducing support for a wide variety of overprinting scenarios and enabling both laser-based (LaserTVAM) and LED-based (LEDTVAM) setups, it offers a versatile and accessible solution for diverse applications. Notably, LEDTVAM setups are not possible to simulate with other existing \added{open-source} software solutions \added{to this extent}.
  
  Here, we demonstrate the capability to directly 3D print inside square perfusable chambers by overprinting around inlets and outlets (\autoref{fig:perfusion}). These experiments highlight the potential impact of our framework in biomedical applications. Notably, square vials are particularly advantageous in biofabrication workflows, as their optically flat surfaces make them directly compatible with confocal and light sheet imaging. Furthermore, our overprinting approach enables the direct fabrication of microfluidic biomimetic channels without the need for additional \removed{post-processing or}assembly steps, thereby minimizing the risk of contamination and mechanical damage. To our knowledge, this is the first time this has been demonstrated with TVAM based methods. \added{There is a similar demonstration by Readily3D SA however it is not a true 3D print and does not overprint occluding inlets.} 
    
    We also demonstrate that, despite the computational load of our framework, 
    we can still generate a robust set of patterns to dynamically adapt to 
    specific situations, such as arbitrarily positioned spheres within resin \autoref{fig:balls}. 
    Sphere detection was accomplished using a simple and cost-effective camera 
    system, while pattern calculation was performed on a standard consumer hardware 
    GPU in less than 30 seconds, all while maintaining a sophisticated optical 
    modeling and optimization scheme. This strategy enables targeted overprinting of specific structures (e.g., \replaced{vascular-like}{3D branching} networks) upon detection of objects inside the printing chamber (e.g., organoids, spheroids, etc.), thereby enhancing precision and adaptability across a wide range of applications, including but not limited to biofabrication.
    This approach is similar to the work by Florczak et al. \cite{Florczak_Groessbacher_Ribezzi_Longoni_Gueye_Grandidier_Malda_Levato_2024}.
    However, our method utilizes a simpler imaging setup to detect the spheres, and we disclose algorithmic details and source code required for fabrication. 
    Furthermore, our entire pipeline — from object detection and pattern optimization to the finished print - is completed in less than three minutes.
    
    By accounting for the light scattering surface 
    of a polished metal rod, we have shown through both experimentation and 
    simulation (\autoref{fig:gears}) that printing fidelity can be improved compared to a simplistic 
    model of a fully light-absorbing rod \cite{kelly2019volumetric}. 
    TVAM patterns are surprisingly robust against simulation-reality mismatch (such as ignoring scattering) but deviations can become problematic. As presented in our experimental findings, fidelity of fine features located closer to the rod degrades from ignoring those effects and more complex light models (such as ours) are required.
    
    \replaced{We further demonstrated}{As last example we demonstrated} the ability to overprint \replaced{engravings}{a symbol} and a lens directly onto a small LED. As shown in \autoref{fig:LED}g, we designed and fabricated an optical system in which any desired pattern can be printed on the LED and projected through a printed lens onto a screen. In this demonstration, we used a $\mathord{\text{\ding{58}}}$ symbol as an example. By incorporating the lens at an appropriate distance from the LED, the system effectively reimages the printed pattern onto the screen, illustrating the potential to integrate arbitrary projection functionalities directly onto compact light sources.

     \removed{As last example, we printed simple lenses onto a glass cuvette (Figure 5). This allows to image samples located inside this glass cuvette. Such overprinting scenarios allow us to fabricate a specialized imaging system where conventional manufacturing methods are more expensive or fail to succeed.}
  
    In summary, Dr.TVAM, our computational framework, has shown to be capable of modeling different optical 
    situations for TVAM. As long as the optical material parameters are known, 
    arbitrary scenarios can be assembled and simulated.

    With our framework we envision many more overprinting scenarios \added{(such as printing on a reflecting mirror as shown in \suppsecnameref{supp:results:dice})}. Since we can model arbitrary shapes and complex material parameters, we are not restricted to the shapes and materials presented here. In fact, the Mitsuba renderer provides a comprehensive library of bidirectional scattering distribution functions (BSDFs), which can be leveraged for accurate modeling of light–surface interactions.
    We believe this significantly expands the application space for TVAM, enabling the creation of complex, functionalized, and multi-component devices from applications in mechanics, optics or biofabrication. \\
    Nevertheless, opportunities for further enhancement of Dr.TVAM remain, including but not limited to modeling of inhibitor diffusion and optical variations caused by polymerization-dependent changes of the refractive index during the printing process

\section*{Methods}
\subsection*{Optical setups}
    \subsubsection*{LaserTVAM}
    \label{sec:lasertvam}
    For the laser-based setup, we utilize blue light with a wavelength of 
    $\lambda=\SI{405}{\nano\meter}$ (HL40033G, Ushio, Japan), which is coupled into 
    a square multimode optical fiber (WF 70×70/115/200/400N, CeramOptec). 
    The projector employed is a high-speed digital micromirror device (DMD) 
    (VIS-7001, Vialux), capable of projecting up to 
    \SI{290}{\hertz} of grayscale images. The light from the blue laser diodes 
    is assumed to be homogeneous and square-shaped after exiting the fiber. 
    The measured maximum continuous power of the light source is approximately 
    \SI{450}{\milli\watt} in the printing plane.
    
    To image the DMD, we use a 4f system consisting of two lenses: a 
    \SI{150}{\milli\meter} lens (LA4874-A-ML, Thorlabs) and a 
    \SI{100}{\milli\meter} plano-convex lens. The Fourier stop 
    is employed to filter out higher diffraction orders. For the rotation stage, 
    we utilize a high-precision rotary stage (X-RSW60C, Zaber). The entire 
    setup is synchronized with electrical output signals from the stage, which are 
    wired through an Arduino Nano Every to the DMD.
    
    Within the vial, the light is assumed to be collimated, allowing for the 
    application of parallel ray optics (as shown in \autoref{fig:perfusion}a). The depth of field is controlled by the 
    diameter of the aperture. After passing through the 4f system, the DMD pixel 
    size is $\SI{20.36}{\micro\meter} \times \SI{20.36}{\micro\meter}$, and the 
    total illuminated area measures $\SI{20.849}{\milli\meter} \times 
    \SI{15.636}{\milli\meter}$ for a pattern resolution of 
    $1024 \times 768$ pixels.
    \added{The depth of field is around \SI{25}{\milli\meter} assuming a Gaussian beam waist half the pixel size in printing space.}

    \subsubsection*{LEDTVAM}
        \label{sec:LEDTVAM}
        The LED-based setup utilizes a light source with a wavelength of 
        $\lambda=\SI{395}{\nano\meter}$ (M395L5, Thorlabs), which has a theoretical 
        maximum power of \SI{1.63}{\watt}. This LED is collimated onto a high-speed 
        digital micromirror device (DMD) (VIS-7001, Vialux) using a condenser 
        lens. The maximum power in the entire image plane is approximately 
        \SI{80}{\milli\watt} for an active pixel area of $700 \times 740$ on the DMD. 
        Due to distortions, we opted to not utilize the entire DMD for projections. 
        
        For imaging the DMD, we employed a single \SI{75}{\milli\meter} lens 
        (AC508-075-A-ML, Thorlabs). A high-precision rotary stage 
        (X-RSW60C, Zaber) is used for rotation. The entire setup is synchronized 
        with electrical output signals from the stage, which are routed through an 
        Arduino Nano Every to the DMD. 
        
\begin{figure}[h!] 
            \includegraphics[width=\textwidth]{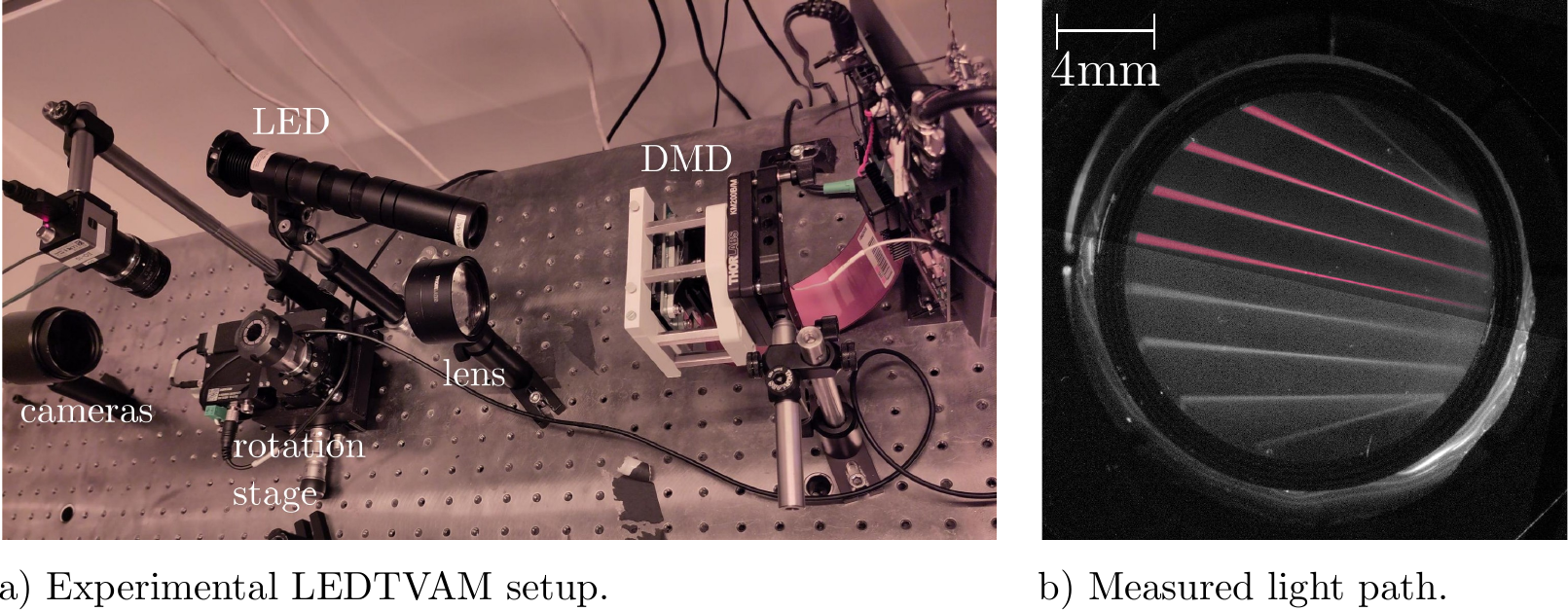}
            \caption{Overview of the LEDTVAM setup.  a) shows the experimental LEDTVAM setup. b) shows a top view of 
            the light path in gray. The pink overlay represents the aligned and 
            simulated light path through a water-filled cylindrical glass vial.}
            \label{fig:LED_setup_combined}
        \end{figure}

        \removed{Since LED-based setups are not yet common in the literature, we aim to provide 
        a more detailed description.} The general configuration can be observed in 
        \autoref{fig:LED_setup_combined}a). The LED was roughly collimated onto the DMD, which is 
        imaged by a single lens. Dr.TVAM is capable of modeling such a setup using three 
        parameters: the field of view in the horizontal direction, the aperture, and 
        the distance between the aperture and the focus point.
        
        We made the light paths visible by adding a fluorescence marker to the water. 
        Based on the experimentally measured light paths in a water-filled vial, as 
        illustrated in \autoref{fig:LED_setup_combined}b), we determined the values of the 
        three parameters. The simulated light path is overlaid in pink. As shown, there 
        is a good match between the experimental results and the simulation.

\subsection*{Vials}
    For the overprinting on metal rods, LEDs, and test tubes we used cylindrical glass vials (Fisherbrand 15 14 0548) whose
    radii are $r_\text{inner} = \SI{6.363 \pm 0.017}{\milli\meter}$ and
    $r_\text{outer}=\SI{7.354 \pm 0.016}{\milli\meter}$. The refractive index is assumed to be $n=1.54$. Adapters to hold the various components in place were designed with CAD software and printed with SLA (Formlabs 3B, Formlabs) using Grey or BioMed black resins (Formlabs).
    As square vials for the perfusion systems we used 25 mm sections of common spectrophotometer cuvettes
    ($L_\text{outer}=\SI{12.43\pm0.03}{\milli\meter}$,
    $L_\text{inner}=\SI{10.07\pm0.03}{\milli\meter}$, $n=1.58$) made of polystyrene
    (7590 30, Brand). Adapters to connect microfluidic tubing and featuring inlet/outlet were designed with CAD software and printed by SLA (Formlabs 3B, Formlabs) using BioMed black resins (Formlabs). We provide access to the design files in \secnameref{sec:codedata}.

\subsection*{Resin preparation}
    \label{sec:app_resin}
    \subsubsection*{Acrylate}
        \label{sec:acrylate_resin}
        
        As acrylate resins (for  
        \suppsecnameref{supp:results:lenses} and \suppsecnameref{supp:results:dice}) we used di-pentaerythritol pentaacrylate (SR399,
        Sartomer, France) which was mixed with \\diphenyl(2,4,6-trimethylbenzoyl)phosphine oxide (TPO, \SI{97}{\percent}; Sigma-Aldrich) in a planetary mixer (KK-250SE,
        Kurabo). The refractive index was experimentally determined to be
        $n=1.4849$.  \\
        As second resin (for \radded{\secnameref{results:gears_rod},} \secnameref{results:led}) we used a mixture of DUDMA (Sigma-Aldrich, USA) and PEGDA 700 (Sigma-Aldrich, USA) \rremoved{and} in a weight ratio of 4 to 1. The refractive index is $n=1.4840$.
        The concentration of TPO varied between \SIrange{10}{50}{\milli\gram}  per \SIrange{40}{70}{\gram} of resin.
        The printing parameters can be seen in \autoref{tab:sim} and for the pattern optimization we measured the resulting absorption coefficients as indicated in \autoref{tab:exp_data_standard}.     
        \added{We do not add any inhibitor except oxygen which is naturally dissolved into the resin during mixing.}
    
    \subsubsection*{Biocompatible resin}
    \label{sec:gelmaresin}
    
    Gelatin-methacryloyl (Gel-MA) was synthesized as previously described \cite{Sgarminato_2024}. The degree of substitution was calculated using 1H-NMR in D2O using internal standard 3-(trimethylsilyl)-1-propanesulfonic acid (DSS, 2H, $\approx\SI{0.75}{ppm}$), and found to be \SI{0.17}{\milli\mol\per\gram}. Gel-MA was dissolved at \SI{37}{\degree} in PBS to result in a \SI{10}{\percent} (w/v) solution. The photoinitiator lithium phenyl-2,4,6-trimethylbenzoylphosphinate (LAP) was added from a 50x stock solution in PBS to obtain a final concentration of \SI{0.05}{\percent} (w/v). The resin was then filter sterilized through \SI{0.2}{\micro\meter} filters. The refractive index was measured to be $n=1.3512$.

\subsection*{Overprinting experiments}

    \subsubsection*{Perfusion system for bio-application\added{s}}
        \label{meth:perfusion}
        After mounting the square vial in its holder, its angular orientation was aligned \added{(accuracy around $\pm\SI{0.5}{\degree}$)} by optimizing the back-reflected signal  of the printing laser. The resulting calibration was then used to initialize another red reference laser to ensure consistent angular registration. Finally, the vial’s vertical position was adjusted under real-time camera observation, enabling precise alignment along the optical axis. \added{We estimate the vertical precision to be around \SI{100}{\micro\meter} with a camera from behind.}
        The straight channels (\autoref{fig:perfusion}g) were designed with a diameter of \SI{0.95}{\milli\meter}. The branched channels (\autoref{fig:perfusion}h) featured a main channel width of \SI{0.95}{\milli\meter} that bifurcated into smaller branches of \SI{0.7}{\milli\meter}. The spiral channel (\autoref{fig:perfusion}i) and all inlets and outlets maintained a tube diameter of \SI{0.95}{\milli\meter}.
    
    \subsubsection*{Context-aware perfusion system}
        \label{sec:sphereover}
        The square cuvettes were prepared by first filling them with a base layer of warm Gel-MA. The resin was allowed to thermally gel in a refrigerator for 15 minutes, after which a glass bead with a diameter of \SI{1.0}{\milli\meter} (Thermo Scientific Chemicals) was placed on the gelled surface. This process was repeated by adding a subsequent layer of resin and a second sphere to embed both objects within the hydrogel matrix.
        
        After placing the square vial into the vial holder, we first aligned its 
        angular orientation. Subsequently, we measured the sphere positions from two 
        views (as illustrated in \autoref{fig:balls}b). This approach enables the 
        determination of the 3D positions in space. Based on these positions, we 
        generated 3D meshes using Python, primarily utilizing \texttt{trimesh}, 
        \texttt{Gmsh}, 
        and \texttt{OpenSCAD}. 
        The generation of the meshes took approximately \SI{10}{\second}. The source code is referenced in \secnameref{sec:codedata}.
        
        Using the generated meshes, we then employed Dr.TVAM to optimize the patterns. 
        We utilized only 100 angular projections and reduced the spatial discretization in object 
        space to $\SI{10.2}{\milli\meter}/128$, which is insufficient to Nyquist sample the resolution of the detector. The optimization took less than \SI{20}{\second}. However, the generated patterns were adequate to produce satisfactory printing results.

    \subsubsection*{Gear on metal rod}
        \label{meth:gears}
        The rough surface of the metal rod was modeled using the Beckmann distribution \cite{Beckmann1987}. \added{The Beckmann distribution describes how strong the light is scattered from the rod's surface (see \secnameref{supp:sec:beckmann})}. The rod's geometry, defined by a mesh file, is modulated with a micro-surface roughness model characterized by the root mean square (RMS) slope. For the metal rod, which had a diameter of \rremoved{$\SI{2.5}{\milli\meter}$}\radded{$\SI{4}{\milli\meter}$}, the roughness parameter was estimated to be \rremoved{$\alpha=0.04$}\radded{$\alpha=0.03$} \radded{in \autoref{fig:gears}b)}.
        \radded{To obtain the experimental optical trace of the scattered light of a single pixel, we filled a square vial with water and a fluorescent dye. This allows us to make the light path visible. The same scenario was simulated in Dr.TVAM and compared to the experimental results.}
        \radded{Lower and higher values of $\alpha$ were tested in \autoref{fig:beckmann2} and show less agreement.}
        The rod was precisely positioned in the center of the vial using a custom SLA-printed adapter.
    
    \subsubsection*{Optical lens with \replaced{engraving}{symbol} on LED}
        \label{meth:LED_over}
        As mentioned, for the \rrreplaced{final two experiments}{last experiment}, LEDTVAM was utilized because it produces smooth, striation-free prints, which enables the fabrication of optical components.
        For the LED overprinting demonstration, commercially available LEDs with a \SI{4.8}{\milli\meter} diameter round cap and a refractive index of $\approx 1.48$ were used. Each LED was secured in the center of the printing vial using a custom SLA-printed holder and submerged directly in the resin. The printed lens was designed with a focal length of \removed{4mm}\added{\SI{4.97}{\milli\meter}}, which is sufficient to project the \replaced{engraved pattern}{symbol} from the LED's surface onto a screen.
        \added{The radius of curvature of the lens surfaces is \SI{4}{\milli\meter}, the experimental measured radius of curvature is \SI{4.08\pm0.05}{\milli\meter}.}

\subsection*{Printing and post-processing}
    \label{sec:app_post}

\subsubsection*{Gel-MA prints}
    First, the chip featuring two SLA printed adapters and a polystyrene cuvette section was assembled via press-fitting. Then, silicone microfluidic tubing (OD:4 mm, ID: 0.8 mm) Saint-Gobain) were connected to the barbed adapters on both sides of the chip. Warm (\SI{37}{\celsius}) Gel-MA photoresin was then injected from one side to fill the chamber. Importantly, the chamber was kept at the right angle to enable evacuation of air bubbles through the venting hole, which was then sealed with a M1 screw. Tubes were then clamped and resin left to thermally crosslink at room temperature for 30 minutes. 
    After printing, the construct was readily placed in a \SI{37}{\celsius} water bath to melt uncrosslinked resin. Warm PBS was then injected via tubing to wash out the channels from residual resin prior to postcuring under UV lamp (Solis-405C, Thorlabs) for 5 seconds. For better visualization, the perfusable constructs were injected with a blue food dye solution in PBS.

    \subsubsection*{Acrylate prints}
        
    Acrylate resin was poured into the glass vials and allowed to sit for a few 
    minutes to remove air bubbles. In cases where small, persistent air bubbles 
    remained, the vials were degassed in an ultrasonic bath for 15 minutes. For 
    various overprinting experiments, the glass vials were equipped with different 
    adapters to accommodate a metal rod, an LED, or a small test tube, as shown 
    in \autoref{fig:gears}, \autoref{fig:LED}, and \autoref{fig:lens_water}. 
    
    After the printing process, the solidified object was placed into a vial 
    containing the solvent propylene glycol monomethyl ether acetate (PGMEA, 
    Sigma-Aldrich). Uncrosslinked resin was washed out by gently shaking 
    the vial with a Reagenzglasschüttler Genie Vortex Mixer Model Vortex-Genie 2 
    for 10 to 25 minutes. The solvent was then replaced with fresh PGMEA, and 
    washing continued for an additional 10 to 25 minutes.
    
    Subsequently, the printed part was submerged in PGMEA containing diphenyl(2,4,6-trimethyl-benzoyl)-phosphine oxide (TPO) for 10 minutes prior 
    to post-curing. The entire vial was then cured under a high-power UV lamp 
    (Solis-405C, Thorlabs) for approximately one minute. 
    \added{For PEDGA/DUDMA resin we replaced PGMEA with \SI{99}{\percent} IPA as the viscosity is lower and IPA was sufficient enough for cleaning.}
    After post-curing, the 
    printed construct was placed on a microscopy coverslip and left to air dry.

\subsection*{Pattern optimization and printing parameters}
    \autoref{tab:sim} provides details regarding the pattern optimization across 
    various experiments. The gradient-based optimization in Dr.TVAM utilized the 
    following loss function (see also \cite{Wechsler_Gigli_Madrid-Wolff_Moser_2024, 
    refId0}):
    
    \begin{align}
        \mathcal{L} &=  \underbrace{w_\text{in} \cdot \sum_{v \,\in\,\text{object}} 
        \text{ReLU}(t_u - I_v)^2}_\text{force polymerization in object} \ \ + 
        \underbrace{w_\text{out}\cdot\sum_{v\,\notin\,\text{object}} 
        \text{ReLU}(I_v - t_l) ^2}_{\text{prevent polymerization elsewhere}} 
        \notag\\[0.2cm]
        &+ \underbrace{w_\text{o} \cdot \sum_{v \,\in\,\text{object}} 
        \text{ReLU}(I_v - 1)^2}_{\text{avoid over-polymerization}} ~\! ~\ \ \ +   
        \underbrace{w_\text{sparsity} \cdot\sum_{j \,\in\,\text{patterns}} 
        |P_j|^D}_{\text{enforce non-sparse patterns}}.
        \label{eq:loss}
    \end{align}
    
    In this equation, $I_v$ represents the absorbed intensity in voxel $v$ after 
    the projection of the patterns. $P_j$ denotes the value of the $j$-th pixel 
    of the patterns. The variables $t_u$ and $t_l$ correspond to the upper and 
    lower thresholds, respectively. The weights $w_\text{in}$, $w_\text{out}$, 
    $w_\text{o}$, and $w_\text{sparsity}$ are relative coefficients for the 
    respective terms. The parameter $D$ imposes varying penalties for sparse 
    values.
    
    Some simulation details are summarized in \autoref{tab:sim}. An online 
    reference for the complete availability of all configuration files, software, 
    and data is provided in \secnameref{sec:codedata}. Most simulations were 
    conducted on a NVIDIA L40S GPU. For the overprinting of the embedded spheres, 
    a desktop computer equipped with a NVIDIA GeForce RTX 4060 Ti was utilized. 
    The spatial discretization and the number of angular patterns for this experiment were 
    significantly reduced, allowing for a substantial decrease in computational 
    time at the expense of quality. Nevertheless, these patterns were adequate to 
    yield satisfactory experimental results.
    
    \begin{table}[h!]
        \centering
        \begin{tabular}{l c c c c c}
    
        {Experiment} &$T_\text{opt}$ in \si{s} & $\eta_\text{energy}$ in \si{\percent}&  
        {$T_L$} & {$T_U$} & $w_\text{sparsity}$  \\
        \hline
        straight channels & 1166 & 1.5  & 0.6 & 0.9 & 0.02 \\
        branched channels & 1175 & 1.5  & 0.6 & 0.9 & 0.02 \\
        helix channels &  1182 & 1.4 & 0.6 & 0.9 & 0.02 \\
        channels connecting spheres &  19 & 6.1 &  0.6 & 0.9 & 0.0002 \\
        absorptive rod & \radded{1460} & \radded{1.6} & 0.7 & 0.92 & 5 \\
        reflective rod & \radded{1937} & \radded{1.6} & 0.7 & 0.92 & 5\\
        \added{lens on LED} & \added{4001} & \added{10.37}  & \added{0.75} & \added{0.93} & \added{0.5}
        \end{tabular}
        \caption{Specifications for the pattern optimization. $T_\text{opt}$ is 
        the time required for the pattern optimization. $\eta_\text{energy}$ 
        indicates the energy efficiency of each set of patterns. {$T_L$} and 
        {$T_U$} are the lower and upper thresholds, respectively, as defined in 
        \autoref{eq:loss}. $w_\text{sparsity}$ governs the sparsity of the 
        patterns.}
        \label{tab:sim}
    \end{table}
    
    Furthermore, the energy efficiency of the patterns varies. It is defined as 
    the ratio of the total laser energy transmitted in one rotation to the total 
    energy that would be transmitted if all patterns were fully activated. 
    Dr.TVAM enables control over the sparsity of the patterns; more sparse 
    patterns result in lower energy efficiency. Conversely, less sparse patterns 
    can influence the histogram while still producing high-fidelity prints. 
    In certain cases, we adjust the parameters to optimize the total printing 
    time. 
    
    It is important to note that, since the loss function is a summation, the 
    sparsity term is contingent upon the configuration and whether LEDTVAM or 
    LaserTVAM is employed. Consequently, the values presented here are not 
    directly interpretable. We just want to indicate that the sparsity term was utilized to 
    fine-tune the efficiency.

\subsection*{Experimental printing parameters}
    
        \autoref{tab:exp_data_standard} presents some of the experimental conditions 
        for the prints. To achieve high-quality prints, we varied the power doses by 
        controlling the current supplied to the laser diodes or the LED.
        
        \begin{table}[h!]
            \begin{tabular}{l c c c c c}
            {Experiment} & setup & {$P$ in \si{\milli\watt\per\milli\meter\squared}} & 
            {$T_\text{print}$ in \si{\second}} & {$\omega$ in \si{\degree\per\second}} & 
            $\mu$ in $\si{\per\milli\meter}$  \\
            \hline
            straight channels & LaserTVAM &$0.77$ & 30.0 & 60.0 & 0.0367\\
            branched channels & LaserTVAM & $0.77$ & 30.0 & 60.0 & 0.0367\\
            helix channels & LaserTVAM &  $0.77$ & 30.6 & 47.0 & 0.0367\\
            channels spheres & LaserTVAM& $0.58$ & 19.2 & 60  & 0.0367\\
            absorptive rod & LaserTVAM & \radded{$0.30$} & \radded{12} & \radded{120} & \radded{0.206} \\
            reflective rod & LaserTVAM & \radded{$0.33$} & \radded{12} & \radded{120} & \radded{0.206} \\
            lenses water & LEDTVAM & $0.33$ & 30.0 & 60.0 & 0.11 \\
            \added{lens LED}  & \added{LEDTVAM} & \added{$0.20$} & \added{25.5} & \added{113} & \added{0.197}\\
            \end{tabular}
            \centering
            \caption{Experimental conditions for the print. $P$ indicates the maximum 
            power per area available. $T_\text{print}$ represents the total printing 
            time. $\omega$ denotes the rotational speed of the stage. $\mu$ is the 
            attenuation coefficient used for absorption.}
            \label{tab:exp_data_standard}
        \end{table}
        
        The printing time for all experiments was well below one minute. Although the 
        number of rotations or the speed was occasionally varied, these factors should 
        not theoretically influence the results. 
        As it can be deduced by their attenuation coefficient $\mu$, for acrylate resins we employed relatively high photoinitiator (TPO) concentrations.
        In contrast, for the bio-resins, we utilized 
        standard  concentrations of the photoinitiator (LAP) commonly used for bioprinting purposes ($\SI{0.05}{\percent}\, \mathrm{w/v}$).

\subsection*{Data availability}
    All raw data generate with Dr.TVAM and the configuration files is available here:\\ \href{https://doi.org/10.5281/zenodo.20038505}{doi.org/10.5281/zenodo.20038505} 

\subsection*{Code availability}
\label{sec:codedata}
    We provide all configuration files for Dr.TVAM and corresponding 3D files for the meshes online to enable reproducibility. Further, we provide access to scripts to generate the context-aware meshes:\\
    \href{https://github.com/EPFL-LAPD/Overprinting-with-Tomographic-Volumetric-Additive-Manufacturing}{github.com/EPFL-LAPD/Overprinting-with-Tomographic-Volumetric-Additive-Manufacturing}\\
   The Zenodo DOI is: \href{https://doi.org/10.5281/zenodo.19187298}{doi.org/10.5281/zenodo.19187298}

\printbibliography
\subsection*{Acknowledgements}
    The authors thank Benoît Vignon and Matthieu Borello for their assistance in designing the SLA-printed adapters and Rami Tabbara and the entire Mitsuba team for helpful discussions regarding Dr.TVAM. We are grateful to Ye Pu for designing the lens for the test tube experiment. We also extend our appreciation to Claude Amendola and the mechanical workshop for the precise cutting of the cuvettes. Finally, special thanks to our lab's \textit{pawfessional}, Lucio, for his support during breaks.\\
This project has received funding from the European Research Council (ERC) under the European Union’s Horizon 2020 research and
innovation program (grant agreement No 948846), and from the Swiss National Science Foundation under project number 196971 - 
\textit{light based volumetric printing in scattering resins}.
It also received funding from the Swiss National Science Foundation 2000-1-240074 under grant number 10007068 - \textit{Neural precision holographic volumetric additive manufacturing}.
\added{R. R. acknowledges Swiss National Science Foundation Return CH Postdoc.Mobility fellowship (P5R5-3\_235066).}

\subsection*{Author contribution}

\begin{tabular}{l l}
    F.W.:&Conceptualization; Investigation; Experiments; Software; Writing – Original Draft.\\
    V.S.:&Conceptualization; Investigation; Experiments; Writing – Review \& Editing.\\
    R.R.:&Conceptualization; Investigation; Experiments; Writing – Review \& Editing.\\
    B.N.:&Formal Analysis; Software; Supervision; Writing – Review \& Editing.\\
    W.J.:&Formal Analysis; Supervision; Writing – Review \& Editing.\\
    C.M.:&Formal Analysis; Supervision; Writing – Review \& Editing.
\end{tabular}

\subsection*{Conflict of interest}
    Christophe Moser is a shareholder of Readily3D SA. All the other authors declare no conflict of interest.

\newpage
\section*{Supplement}
\setcounter{figure}{0}
\renewcommand{\thefigure}{S\arabic{figure}}
\setcounter{table}{0}
\renewcommand{\thetable}{S\arabic{table}}
\appendix
\section{Lenses on test tube}
          \label{supp:results:lenses}
         
         \added{In this\removed{last} experiment, we demonstrate printing a lens onto a cylindrical glass tube filled with water.
            The lens is designed to image samples immersed in the water inside the tube.}
        \begin{figure}[h]
            \centering
            \includegraphics[width=\textwidth]{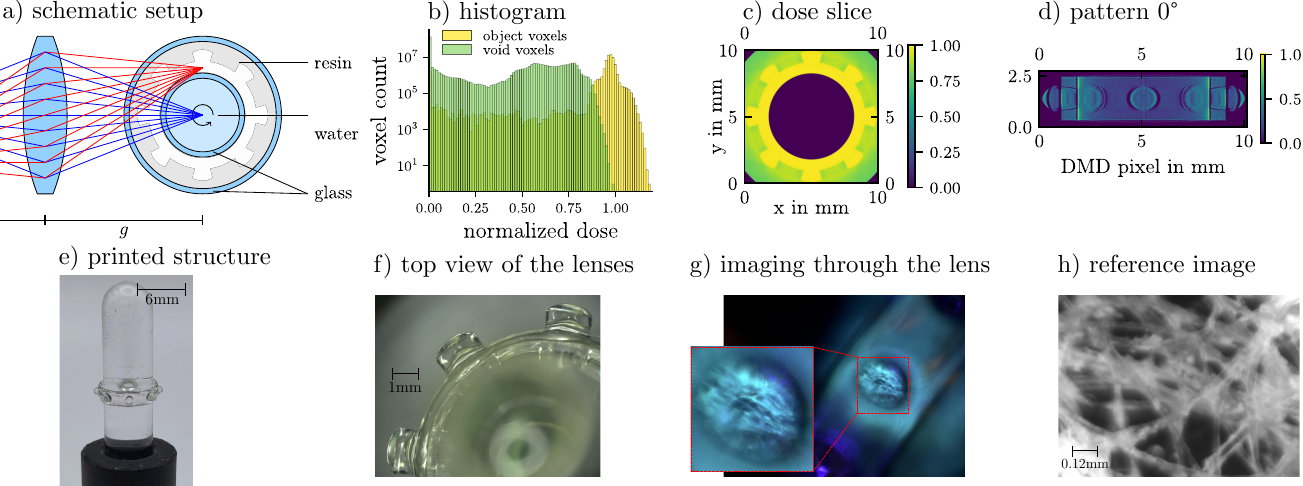}
                \caption{\added{Printing a lens on a water-filled glass cuvette for in-situ imaging. 
                a) Schematic of the experimental setup. 
                b) Voxel \replaced{intensity}{dose} histogram from the simulation. 
                c) A slice of the cumulative \replaced{intensity}{dose} map. 
                d) Example projection patterns from two different angles. 
                e) The final printed lens on the cuvette. 
                f) Top view of the lenses on the cuvette.
                g) Illustration of the lens's intended function for imaging a sample within the tube.
                h) A reference image of the same sample.}}
            \label{fig:lens_water}
        \end{figure}
\added{
         As in the previous examples, the optical paths are complex due to the multiple media involved in the light propagation. The embedded water, in particular, has negligible absorption but a refractive index that differs significantly from the other materials. Therefore, it is crucial to account for the multiple refractions occurring at the interfaces between vial, resin, small vial, and water.\\
         A schematic of the setup is shown in \autoref{fig:lens_water}a). Optimizing the patterns for this geometry yields the histogram in \autoref{fig:lens_water}b)  and a \replaced{intensity}{dose} slice in \autoref{fig:lens_water}c). Some object voxels fall within the water-filled cuvette; while these are not printed, they contribute to the apparent discrepancies in the histogram. One example pattern is shown in \autoref{fig:lens_water}d).\\
        The final printed lens structure on the inner glass cuvette is shown in \autoref{fig:lens_water}e), with a detailed view in \autoref{fig:lens_water}f). The lens is designed to image samples placed inside the water-filled tube.
         As a sample we used optical tissue paper, which was stained with text marker. We subsequently used its qualitative working principle \replaced{is}{as} shown in \autoref{fig:lens_water}g). 
        A similar reference image of the sample is presented with a $4 \times$ microscope in \autoref{fig:lens_water}h). The contrast is low because the tissue was embedded in water and the fluorescent marker diffused into the water and contributes to background light.
        \label{meth:lensestest}
        A test tube with an inner radius of $r_\text{inner} = \SI{2.58}{\milli\meter}$, an outer radius of $r_\text{outer} = \SI{3.18}{\milli\meter}$, and a refractive index of $1.54$ was used as the substrate to print on. The tube was filled with water and secured in the center of the larger printing vial using an SLA-printed cap. The microlenses were designed to image the central axis of the test tube.  
}

\section{\added{Dice on mirror}}
          \label{supp:results:dice}

\begin{figure}[h]
            \centering
            \includegraphics[width=\textwidth]{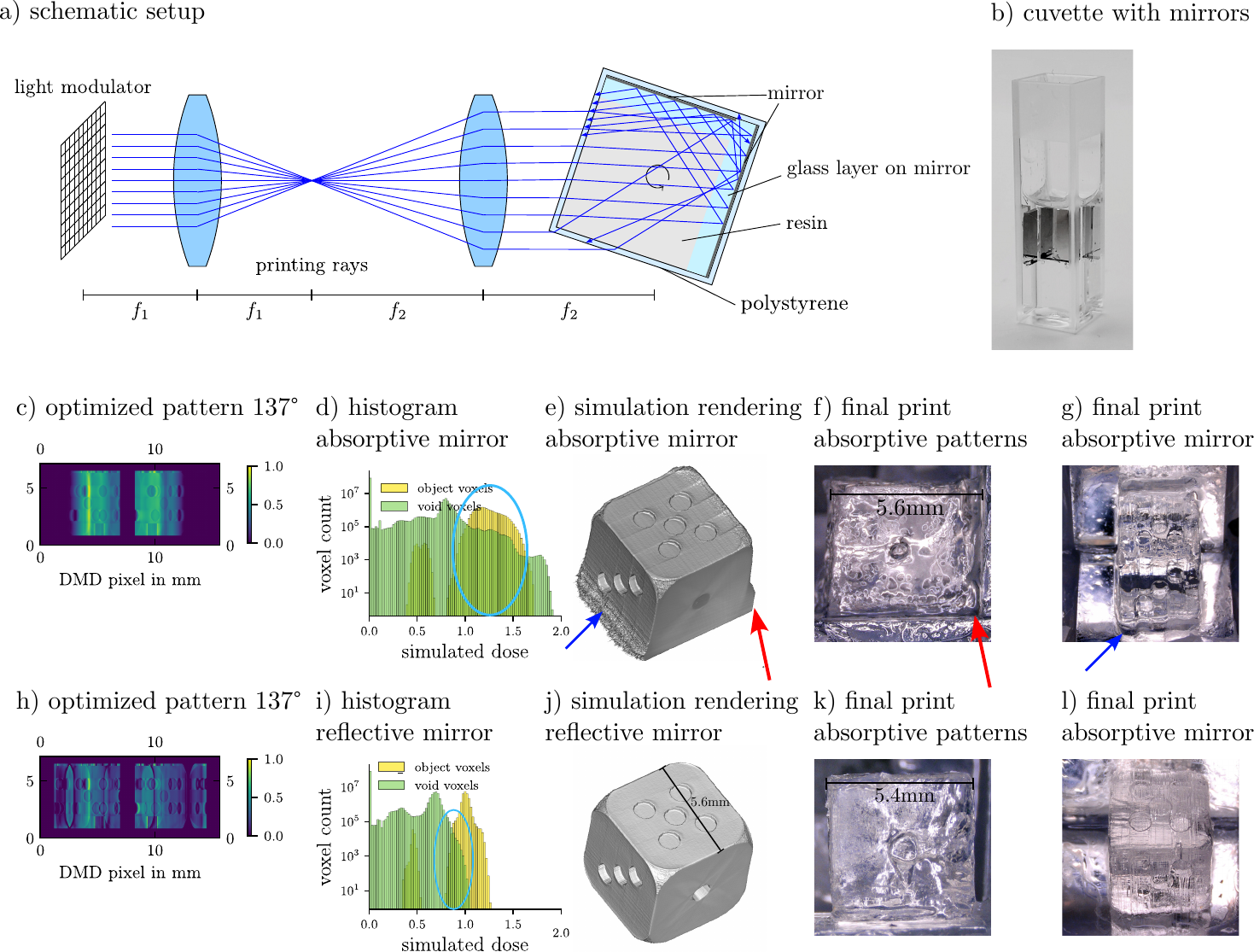}
                \caption{\added{Overprinting of a dice onto a reflective mirro surface. a) schematic experimental setup. b) two mirrors glued into a quadratically shaped cuvette. c) optimized patterns for absorbing mirrors. d) \replaced{intensity}{dose} histogram for patterns with absorbing mirrors projected onto reflecting mirrors. e) best simulation rendering for that case. f) experimental print of absorbing patterns on reflective mirror. g) experimental print of absorbing patterns on reflective mirror. h) optimized patterns for reflective mirrors. i) \replaced{intensity}{dose} histogram for patterns with reflective mirrors. j) best simulation rendering for that case. k) experimental print of reflective patterns. l) experimental print of reflective patterns.}}
            \label{fig:mirror}
        \end{figure}

\added{In this section we demonstrate printing onto a smooth mirrored surface. 
The general setup for this print is shown in \autoref{fig:mirror}a). Because of the internal reflection the ray-path is complex as qualitatively shown. 
We fabricated the cuvettes by gluing custom cut mirrors on the inside of the square cuvette. An example cuvette filled with resin is displayed in \autoref{fig:mirror}b).
This arrangement implies that there is a glass layer on top of the mirror. We model all interfaces and reflection with our software Dr.TVAM.
}

\added{
To showcase our software, we compare two scenarios. The first scenario assumes that the mirror is fully light absorbing. One specific pattern of the optimization is in \autoref{fig:mirror}c). 
If we project those patterns, we already identify a certain overlap in the histogram in \autoref{fig:mirror}d) (marked by the blue ellipsis). Also, the 3D rendering of the best possible theoretical print in \autoref{fig:mirror}e) shows some deformations from an ideal cube. Further, our experimental prints with the absorptive patterns show some deformations in \autoref{fig:mirror}f) and g). See the red arrow for the most severe discrepancies. The gap between one of the mirrors and the cube is closed, even though by design it should not be. There is also a edge artefact on one side (blue arrow).\\
In our framework we can correctly simulate patterns with a large amount of reflection, as seen in \autoref{fig:mirror}h) where multiple reflections of the mirrors can be observed. The histogram in \autoref{fig:mirror}i) displays higher fidelity. The final rendering of the reflection aware patterns is in \autoref{fig:mirror}j). This time, the overall quality is slightly higher and the gap between the second mirror is preserved.
The final, well preserved prints are shown in \autoref{fig:mirror}k) and l).
}

\section{\added{Rendering of the channels}}
\begin{figure}[h]
            \centering
            \includegraphics[width=.7\textwidth]{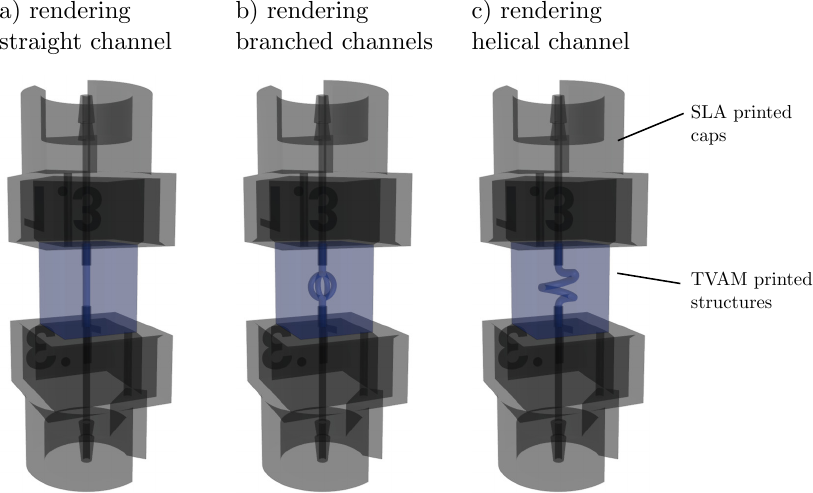}
                \caption{\added{Rendering of the TVAM printed channels with SLA printed caps. a) straight channel, b) branched channels and c) helical channel.}}
            \label{fig:rendering_channels}
        \end{figure}

    \added{\autoref{fig:rendering_channels} contains a Blender rendering of the printed channels of the perfusable chips.
    The black caps at the top and bottom were SLA fabricated and contain the nozzle inlets.}

    \added{Note, we did not render the square cuvette itself since this is not modelled with a mesh file but instead is handled by our software Dr.TVAM.}

\section{Exposure series}
\label{supp:sec:exposure_series}
\radded{To rule out the possibility that the visible defects in \autoref{fig:gears} for the gear printed from the uncorrected pattern are caused by operator-induced under- or overexposure, we present an exposure series in \autoref{fig:exposure_series}. The top row shows prints obtained at different exposure powers using the uncorrected pattern, whereas the bottom row shows the corresponding results for the scattering-corrected pattern.}

\radded{The fine features located close to the rod are consistently overpolymerized in the prints produced from the uncorrected pattern. Notably, even under substantial underexposure, overpolymerization persists near the rod, while the remainder of the gear is clearly underexposed.}

\radded{In contrast, the holes are generally much better preserved for the scattering-corrected patterns. As this behavior is also in good agreement with the simulations, we can exclude the existence of an exposure power for the uncorrected pattern that would produce a print with well-preserved holes.}

\begin{figure}[h]
    \centering
    \includegraphics[width=\textwidth]{figures/gear_exposure_series.pdf}
    \caption{\radded{Exposure series for the gear structures. Top row: a)--d) prints obtained from the uncorrected pattern at different exposure powers. Bottom row: e)--h) corresponding prints obtained from the scattering-corrected pattern.}}
    \label{fig:exposure_series}
\end{figure}

\section{Beckmann distribution to model a scattering rod}
\label{supp:sec:beckmann}
\added{The Beckmann distribution is a mathematical model that describes how light scatters from rough surfaces by characterizing the statistical distribution of microscopic surface normals (microfacets). In computer graphics rendering systems like Mitsuba (which Dr.TVAM is based on), it serves as a physically-based microfacet normal distribution function (NDF) derived from Gaussian random surfaces, enabling realistic simulation of surface roughness and light reflection.\\
The Beckmann distribution is characterized by a single parameter $\alpha$ which describes how strong the light is scattered.}
\rremoved{For our experiments in  we assumed a value of $\alpha=0.04$ which was visually matched to comparison renderings.}

\begin{figure}[h]
    \centering
    \includegraphics[width=\textwidth]{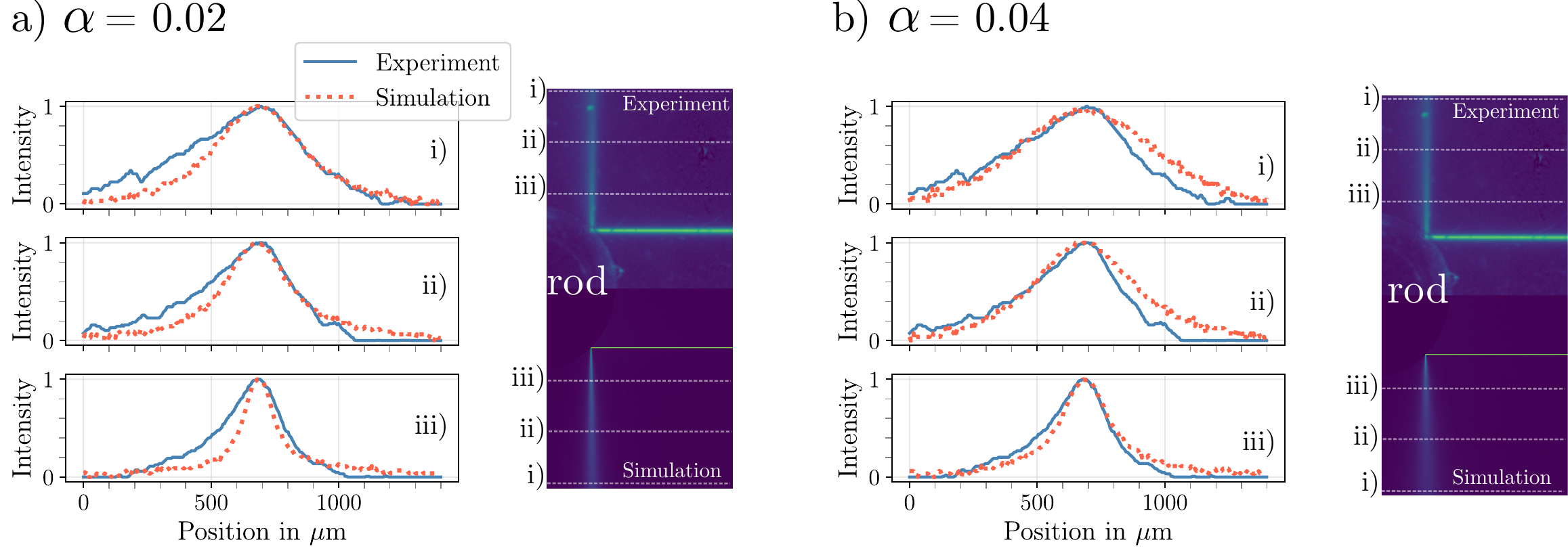}
    \caption{\radded{Characterization of the $\alpha$ parameter. The experimental captured image and the simulated response and three cross section at different distances to the rod. 
    a) shows the results for $\alpha=0.02$ and b) for $\alpha=0.04$.}}
    \label{fig:beckmann2}
\end{figure}

\radded{In our experiment, we estimated $\alpha$ by projecting a single pixel onto the metal rod. We embedded the rod inside a fluorescent water bath. This allows us to make the scattering light visible with a camera from below. From the spread of the reflected light we can then estimate $\alpha=0.03$.
\autoref{fig:beckmann2} shows the reflected light curves for $\alpha=0.02$ and $\alpha=0.04$ which show a good, but not a perfect fit. The results indicated in \autoref{fig:gears}b) with $\alpha=0.03$ show a much better fit.}

\begin{figure}[H]
    \centering
    \includegraphics[width=\textwidth]{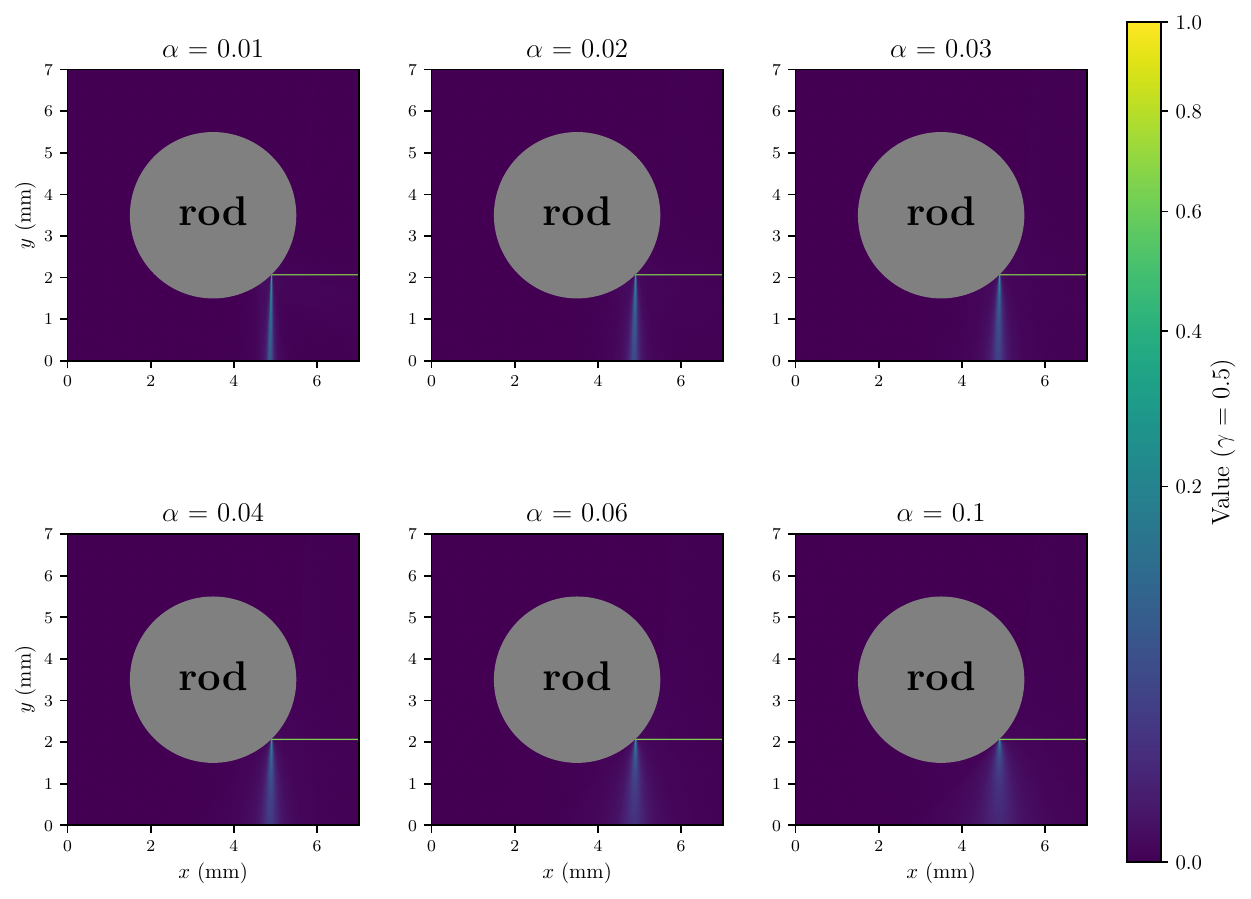}
        \caption{The deposit intensity from a single pixel illuminating the rod from the right which results in scattered paths.
        We show different values of $\alpha$ used for the Beckmann distribution from the metal rod. Note, we display the intensity applied with a gamma factor of $\gamma=0.5$ to make the weakly scattered rays visible. \rremoved{We display the 2D scattering on a rod, in our simulation we use the real 3D scattering.}}
    \label{fig:beckmann}
\end{figure}

\added{In \autoref{fig:beckmann} we illustrate how a single pixel projects light (from right to left) onto a metal rod with varying values of $\alpha$. The light intensity is recorded according to the absorbing medium. To make the low intensity of the scattered light visible, we show the image with a gamma correction of \radded{$\gamma=0.5$}. As it can be seen, higher values of $\alpha$ indeed scatter the light stronger.}

\section{\added{Dose histograms}}
\begin{figure}[h]
            \centering
            \includegraphics[width=1\textwidth]{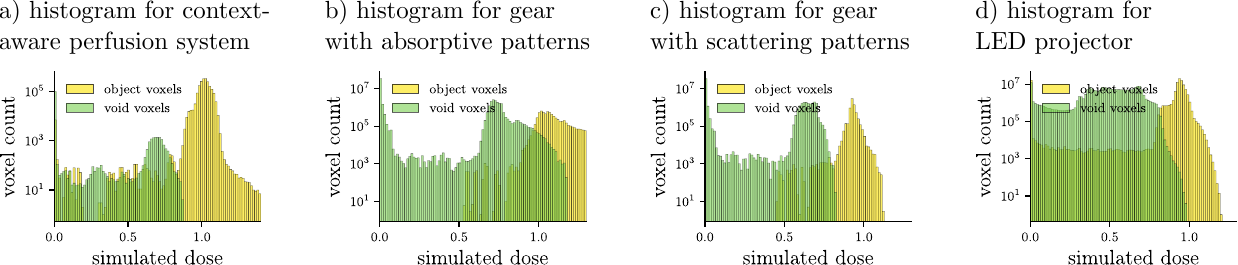}
                \caption{\added{Simulated dose histograms for a) the context-aware perfusion system, b) the absorptive patterns for the gear applied to a scattering gear overprint, c) the scattering patterns of the gear overprinting and d) for the LED overprinting}.}
            \label{fig:all_histograms}
        \end{figure}

    \added{\autoref{fig:all_histograms} displays the dose histograms for the experiments in  \secnameref{results:channels}, \secnameref{results:gears_rod} and \secnameref{results:led}. Dose histograms are an effective way to judge the quality of the projection patterns. Note, all histograms are in a logarithmic scale.
    For example, \autoref{fig:all_histograms}a), c) and d) results in high quality prints as there is no significant overlap.
    However, \autoref{fig:all_histograms}b) shows a noticeable overlap as we apply patterns simulated in an absorptive gear case to a scattering gear.}
\added{Notice, the yellow dose in \autoref{fig:all_histograms}d) corresponds to voxels partially inside the LED hence they do not result in visible quality loss.}

\section{\added{Printing Parameters}}
\added{This section contains the printing and optimization parameters for the experiments presented in \suppsecnameref{supp:results:lenses} and \suppsecnameref{supp:results:dice}}.
   \begin{table}[h!]
        \centering
        \begin{tabular}{l c c c c c}
    
        {Experiment} &$T_\text{opt}$ in \si{s} & $\eta_\text{energy}$ in \si{\percent}&  
        {$T_L$} & {$T_U$} & $w_\text{sparsity}$  \\
        \hline
        {lenses for imaging} & {4019} & {12.3} & {0.8} & {0.95} & $10^{-25}$ \\
        absorptive mirror &  1107 & 0.07 &0.7 & 0.94 & 2\\
        reflective mirror &   1741 & 0.01 & 0.7 & 0.94 & 2
        \end{tabular}
        \caption{\added{Specifications for the pattern optimization. $T_\text{opt}$ is 
        the time required for the pattern optimization. $\eta_\text{energy}$ 
        indicates the energy efficiency of each set of patterns. {$T_L$} and 
        {$T_U$} are the lower and upper thresholds, respectively, as defined in 
        \autoref{eq:loss}. $w_\text{sparsity}$ governs the sparsity of the 
        patterns.}}
        \label{tab:sim2}
    \end{table}

        \begin{table}[h!]
            \begin{tabular}{l c c c c c}
            {Experiment} & setup & {$P$ in \si{\milli\watt\per\milli\meter\squared}} & 
            {$T_\text{print}$ in \si{\second}} & {$\omega$ in \si{\degree\per\second}} & 
            $\mu$ in $\si{\per\milli\meter}$  \\
            \hline
            lenses for imaging & LEDTVAM & $0.33$ & 30.0 & 60.0 & 0.11 \\
            absorptive mirror & LaserTVAM & 1.22 & 25.2 & 100 & 0.129\\
            reflective mirror & LaserTVAM & 0.95 & 28.8 & 100 & 0.129\\
            \end{tabular}
            \centering
            \caption{\added{Experimental conditions for the print. $P$ indicates the maximum 
            power per area available. $T_\text{print}$ represents the total printing 
            time. $\omega$ denotes the rotational speed of the stage. $\mu$ is the 
            attenuation coefficient used for absorption.}}
            \label{tab:exp_data_standard2}
        \end{table}

\end{document}